\documentstyle[12pt,psfig]{article}
\begin{document}
\begin{titlepage}
\begin{center}

{\Large\bf{The Asymptotic Regime of High Density QCD}}
\\[5.0ex]
{\Large\it{ M. B. Gay  Ducati $^{*}$\footnotetext{$^{*}$E-mail:gay@if.ufrgs.br}}}
 {\it and}
{ \Large \it{ V. P.  Gon\c{c}alves $^{**}$\footnotetext{$^{**}$E-mail:barros@if.ufrgs.br} 
}} \\[1.5ex]
{\it Instituto de F\'{\i}sica, Univ. Federal do Rio Grande do Sul}\\
{\it Caixa Postal 15051, 91501-970 Porto Alegre, RS, BRAZIL}\\[5.0ex]
\end{center}

{\large \bf Abstract:}
We discuss the distinct approaches for high density QCD (hdQCD) in the asymptotic regime of large values of parton  density. We derive the AGL equation for running coupling constant and obtain the asymptotic solution, demonstrating  that the property of partial saturation of the solution of the AGL equation is not modified by the running of the coupling constant. We show that in this kinematical regime, the solution of the AGL equation coincides with the solution of an evolution equation, obtained recently using the McLerran-Venugopalan  approach. Using the asymptotic behavior of the gluon distribution we calculate the $F_2$ structure function assuming  first that the leading twist relation between these two quantities is valid and second that this relation is modified by the higher twist terms associated to the unitarity corrections. In the first case we obtain that the corresponding $F_2$ structure function  is linearly proportional to $ln \, s$, which agrees with the results obtained recently  by Kovchegov using a distinct approach. In the second case a softer behavior is obtained. In both cases, the $F_2$ structure function unitarizes and the Froissart boundary is not violated in the asymptotic regime of high density QCD. We conclude that the partial saturation of the gluon distribution and the unitarization of the structure  function   are general results, well-established by three distinct approaches for high dense systems.

\vspace{1.5cm}



\end{titlepage}

\section{Introduction}

The parton high density regime in deep inelastic scattering is one of the  frontiers in perturbative QCD (pQCD). This corresponds to the small $x$ region and represents the challenge of studying the interface between the perturbative and nonperturbative QCD, with the peculiar feature that this transition is taken in a kinematical region where the strong coupling constant $\alpha_s$ is small. By the  domain of perturbative QCD we mean the region where the parton picture has been developed and the well-established methods of operator expansion and renormalization group equations have been applied sucessfully. The Dokshitzer-Gribov-Lipatov-Altarelli-Parisi (DGLAP) equations \cite{dglap}, which are based upon the sum of QCD ladder diagrams, are the evolution equations in this kinematical region. In the limit of small values of $x$ $( < 10^{-2})$, on the other hand, one expects to see new features inside the nucleon: the density of gluons and quarks becomes very high and a new dynamical effect associated with the unitarity corrections is expected to stop the further growth of the structure functions. Ultimately, the physics in the region of high parton densities will be described by nonperturbative methods, which is still waiting for a satisfactory solution in QCD. However, the transition from the moderate $x$ region towards  the small $x$ limit may be very well be accessible in perturbation theory, and, hence, allows us to test the ideas about the onset of nonperturbative dynamics. In the QCD at dense systems (hdQCD) we still have a small parameter (the coupling constant $\alpha_s$), which implies that  we can start to approach this regime using the developed methods of perturbative QCD.

The expectation of the transition for the high density regime can be understood  considering  the physical picture of the deep inelastic scattering. In the infinite momentum frame (IMF) the virtual photon with virtuality $Q^2$ measures the number density of charged     partons having longitudinal momentum fraction $x$ and transverse spatial size $\Delta x_t \le 1/Q$. When $Q^2$ is large, $\alpha_s(Q^2)$ is small, so that the struck quark can be treated perturbatively. Also, when $Q^2$ is large the struck quark is small, $\Delta x_t \approx 1/Q$, so that one can picture the struck quark as being isolated, far away from similar quarks, in the proton. Thus, so long as the parton distributions are not large, the partons in a proton are dilute. However, if the parton distributions get large enough, which happens when $x$ is very small, partons in the proton must begin to overlap. If there is a sufficient amount of parton overlap then a given parton will not act as a free quantum over its lifetime but will interact strongly with the other partons in the proton, even though $\alpha_s$ may still be in the perturbative regime.

The behavior of the cross sections in the high energy limit ($s \rightarrow \infty$) and fixed momentum transfer is expected to be described by the BFKL equation \cite{bfkl}. The simplest process where this equation applies is the high energy scattering between two heavy  quark-antiquark states, {\it i.e.} the onium-onium scattering. This process was studied in the dipole picture \cite{muedip}, where the heavy quark-antiquark pair and the soft gluons  in the limit of large number of colors are viewed as a collection of color dipoles. However, one of the main characteristics of the BFKL equation is that it predicts very high density of partons in the small $x$ region. The power behavior of BFKL cross section violates the Froissart boundary, which implies that the BFKL Pomeron describes only the pre-asymptotic behavior at not very  large energies and in order to find the true high energy asymptotics in perturbative QCD we need to unitarize the BFKL Pomeron.  The understanding and analytical description of the unitarity corrections and consequently of the high density QCD (hdQCD) is currently one  great challenge.

About 17 years ago, Gribov, Levin and Ryskin (GLR) \cite{100} have started the description of the high density systems in the double logarithmic approximation (DLA) of perturbative QCD. They argued that the physical process of  recombination of partons become important in the parton cascade at  a large value of the parton density, and that these unitarity corrections could be expressed in a new evolution equation - the GLR equation. This equation considers the leading nonladder contributions: the multiladder diagrams, denoted as fan diagrams. The main characteristics of this equation are that it truncates the series in the expansion in powers of the density at the first nonlinear term  and predicts the saturation of the gluon distribution at very small $x$ \footnote{We should be careful with this statement, since it is possible that the region of validity of the GLR equation ends before saturation is reached \cite{bartels1}.}; it predicts a critical line, separating the perturbative regime from the saturation regime; and it is only valid in the border of this critical line. Therefore, the GLR equation predicts the limit of its validity, and as all truncation is valid in a limited range. In the last decade, the solution \cite{collins,bartels1,bartels2} and possible generalizations \cite{bartels,laenen} of the GLR equation  have been studied in great detail. At the  moment, there are basically three distinct approaches for the dynamics at high densities:
\begin{itemize}
\item the Ayala, Gay Ducati and Levin (AGL) approach \cite{ayala1}, which considers the multiple pomeron exchange in the DLA limit of perturbative QCD, considering as basic degree of freedom the usual partons (quarks and gluons). The starting point of this approach is the proof of the Glauber formula in QCD \cite{muegla}, which considers only the interactions of the fastest partons with the target. In \cite{ayala1} the authors have proposed a generalized evolution equation (the AGL equation) which takes into account the interaction of all partons in the partons cascade with the target  and have analysed the solutions of this equation considering the semiclasssical method. The main conclusion was  that the unitarity corrections strongly modify the behavior of the gluon distribution in the small $x$ region;
\item the McLerran-Venugopalan {\it et al.} (MV-JKLW) approach \cite{mv,jamal}, which is based on the effective Lagrangian formalism for the low $x$ DIS and the Wilson renormalization group. The basic degree of freedom is the gluonic field. In \cite{jamal} the authors have derived a general evolution equation for the gluon distribution in the limit of large parton densities and leading logarithmic approximation considering a very large nucleus. In the general case this evolution equation is a functional equation, which does not allow to obtain analytical solutions. Recently, these authors have considered the DLA limit on their result \cite{jamal2} and have shown that the evolution equation reduces to  an equation with a functional form similar, but not identical, to the AGL equation. The solution of  this equation was discussed in Ref. \cite{jamalwang} considering the semiclassical method;
\item the Kovchegov (K) approach \cite{kov}, which is based in the resumation of the multiple pomeron exchange in the leading logarithmic $1/x$ approximation [LLA($1/x$)] using the dipole picture, where the basic degree of freedom are quark-antiquark dipoles. The author has obtained an evolution equation for the interaction cross section of the $q \overline{q}$ pair with  a nucleus considering the multiple scattering of the dipoles, which unitarizes the BFKL Pomeron. The solution of this evolution equation was analysed by the author in Ref. \cite{kov2} and independently in Ref. \cite{levtu}.
\end{itemize}
In contrast with the MV-JKLW and K approaches, in the AGL approach a comprehensive phenomenological analysis of the $ep$ \cite{ayala3,vic1,vic2} and  $eA$ \cite{prc,df2a} processes exists. The analysis of the behavior of the distinct observables for the HERA kinematical region using the Glauber-Mueller formula was presented in Ref. \cite{ayala3}. In this kinematical region the solutions from the AGL equation and the Glauber-Mueller formula approximately coincide \cite{ayala1}. The results from these analysis agree with the recent HERA data and allows to make some predictions which will be tested in a near future. Our main conclusion was that the unitarity corrections cannot be disregarded in the HERA kinematical region.

All these approaches reproduce the small-$x$  limit of the DGLAP evolution equations in the DLA limit of pQCD and the GLR equation as a first order unitarity correction. However, important questions still  need to be answered: Since at high densities higher orders cannot be disregarded, what is the correct framework to treat these unitarity corrections ?  Is there a common limit between these approaches ?

Although the complete demonstration of the equivalence between these three approaches is still an open question, a first step was given recently in the Ref. \cite{npbvic}. In that paper we  demonstrated the equivalence between the K equation and the AGL equation in the DLA limit, which motivates the use of the AGL approach, which considers the medium effects associated with the high density present at small values of $x$ and/or nuclear collisions at high energies, as the dynamics at dense systems. Our goal in this paper is to give a second step in the demonstration of the equivalence between the three approaches. Since in the general case the comparison between these approaches is a very difficult task, we will consider the asymptotic regime of very small values of $x$ (very high densities) and demonstrate that in this regime there is an agreement in the predictions of these approaches, establishing a general behavior for the $F_2$ structure function in this kinematical region, which does not violate the Froissart boundary.

The paper is organized as follows: In the next section we briefly review the derivation of the AGL equation and the asymptotic solution for fixed $\alpha_s$ in order to fix notation and to present known results \cite{ayala1}. Later we derive the AGL equation for running $\alpha_s$ and consider the asymptotic solution of this equation. We demonstrate that the main feature of the solution of the AGL equation, the partial saturation of the gluon distribution, is not modified by the inclusion of the running of the coupling constant. In the last subsection we assume the evolution equation proposed by Jamal Jalilian Marian {\it et al.} in Ref. \cite{jamal2} and consider its asymptotic limit of high densities, showing that in this limit the predictions for the gluon distribution of this approach coincides with the AGL result. We will assume then the universality of the $x$ and $Q^2$ dependence of the gluon distribution (the partial saturation) in the asymptotic regime of high density QCD and calculate the asymptotic behavior of the $F_2$ structure function in Section 3. We will consider two possibilities to calculate the structure function using as input the gluon distribution. First, we assume that the  leading twist relation between $F_2$ and $xG$ is not modified by the unitarity corrections. In this case we demonstrate that our results coincide with the result obtained independently by Kovchegov in Ref. \cite{kov2}, and that $F_2$ does not violate the Froissart boundary in the asymptotic regime. This is a general behavior predicted by all approaches using the leading twist relation. Second, we consider that the relation between $F_2$ and $xG$ is modified by the higher twist terms associated with the unitarity corrections. We demonstrate that the behavior      of the structure function is modified by the presence of these terms, but the    Froissart boundary is not violated. Finally, in Section 4 we present our conclusions. 

We have tried to make this paper self-contained, and therefore have included some of the material already included in the early works \cite{ayala1,jamal2}.

\section{The AGL Equation}

\subsection{Derivation and Asymptotic Solution for fixed $\alpha_s$}

We start considering  the interaction between a virtual colorless hard probe and the nucleus via a gluon pair component of the virtual probe.  The cross section for this process is written as
\begin{eqnarray}
\sigma^{G^*A}=  \int_0^1 dz \int \frac{d^2r_t}{\pi} 
 |\Psi_t^{G^*}(Q^2,r_t,x,z)|^2 \sigma^{gg+A}(z,r_t^2)\,\,,
\label{sig1}
\end{eqnarray}
where $G^*$ is the  virtual colorless hard probe with virtuality  $Q^2$, $r_t$ is the transverse separation of the pair,  $z$ is the fraction of energy carried by the gluon and $\Psi_t^{G^*}$ is the wave function of the transverse polarized gluon in the virtual probe. Furthermore, $\sigma^{gg+A}(z,r_t^2)$ is the cross section of the interaction of the $gg$ pair with the  nucleus.
To estimate the unitarity corrections we have to take into account the rescatterings of the gluon pair inside the nucleus. Using the Glauber-Mueller approach, which  considers the interations of the fastest partons with the target, we get 
\begin{eqnarray}
\sigma^{G^*A}=  \int_0^1 dz \int \frac{d^2r_t}{\pi} 
\int \frac{d^2b_t}{\pi} |\Psi_t^{G^*}(Q^2,r_t,z)|^2 \,2\,
[1 - e^{-\frac{1}{2}\sigma_N^{gg}(x^{\prime}
,\frac{4}{r_t^2})S(b_t)}]\,\,, 
\label{sig3}
\end{eqnarray}
where $x^{\prime} = x/(z\,r_t^2\,Q^2)$ ($x$ is the Bjorken variable), $ b_t$ is the impact parameter,  
$S(b_t) = (A/ \pi R_A^2) e^{-\frac{b_t^2}{R_A^2}}$ is the gaussian profile 
function  and $\sigma_N^{gg} = \frac{C_A}{C_F} \sigma_N^{q\overline{q}}$ is the cross section of the 
interaction of the $gg$ pair with the  nucleons inside  the 
nucleus. It was   shown \cite{plb} that
\begin{eqnarray}
  \sigma_N^{q\overline{q}} 
= \frac{C_F}{C_A} (3 \alpha_s(\frac{4}{r_t^2})/4)\,\pi^2\,r_t^2\,
 xG_N(x,\frac{4}{r_t^2}) \,\,,
\label{sigqq}
\end{eqnarray}
 where  $xG_N(x,\frac{4}{r_t^2})$ is the nucleon gluon   distribution.

The  relation $\sigma^{G^*A}(x,Q^2) = (4\pi^2 \alpha_s/Q^2)xG_A(x,Q^2)$ is valid for a virtual probe $G^*$ with virtuality $Q^2$.  
Consequently, using  the expression of the squared  wavefunction 
 we obtain that   
the Glauber-Mueller formula for the interaction of the $gg$ pair with the 
nucleus is written as
\begin{eqnarray}
xG_A(x,Q^2) = \frac{4}{\pi^2}   \int_x^1 
\frac{dx^{\prime}}{x^{\prime}} \int_{\frac{4}{Q^2}}^{\infty} \frac{d^2r_t}{\pi r_t^4}
\int \frac{d^2b_t}{\pi} \,2\, [1 - e^{- \frac{1}{2} \sigma_N^{gg} (x^{\prime}, \frac{4}{r_t^2}) S(b_t)}]\,\,. 
\label{gm}
\end{eqnarray}

The AGL equation can be obtained directly from the above equation  
differentiating this formula with respect to $y = ln \,1/x$ and 
$\epsilon = ln \, Q^2/\Lambda_{QCD}^2$. Therefore the AGL equation is given by 
\begin{eqnarray}
\frac{\partial^2 xG_A(x,Q^2)}{\partial y \partial \epsilon} =
\frac{2\,Q^2}{\pi^2} \int \frac{d^2b_t}{\pi}  \,[1 - e^{-\frac{1}{2}\sigma_N^{gg}(x
,Q^2)S(b_t)}]\,\,, 
\label{agl}
\end{eqnarray}
where the dependence of  $\sigma_N^{gg}$ in the  virtuality of the   
virtual probe results from the derivative. The nonperturbative effects 
coming from the large distances are absorbed in the boundary and initial 
conditions. This equation is valid in the double logarithmic approximation (DLA).

In our calculations we use the  Gaussian parameterization for the  profile function, which implies that the integration over $b_t$ is straithforward. Moreover, in what follows we will consider the nucleon case ($A = 1$). Then, we get
\begin{eqnarray}
\frac{\partial^2 xG(x,Q^2)}{\partial y \partial \epsilon} =
\frac{2\,Q^2 R^2}{\pi^2} \{ C + ln [\kappa_G (x,Q^2)] + E_1 [\kappa_G (x,Q^2)]\} \,\,,
\label{agl2}
\end{eqnarray}
where $C$ is the Euler constant, $E_1$ is the exponential function and the function $\kappa_G$ is defined by
\begin{eqnarray}
\kappa_G (x,Q^2) \equiv \frac{\alpha_s N_c  \pi }{2 Q^2 R^2} xG(x,Q^2)\,\,,
\label{kapag}
\end{eqnarray}
and represents the probability  of gluon-gluon interaction inside the parton cascade.
Using the above definition for $\kappa_G$ we can rewrite the expression (\ref{agl2}) in a more convenient form (for fixed $\alpha_s$)
\begin{eqnarray}
\frac{\partial^2 k_G (y,\epsilon)}{\partial y \partial \epsilon} + \frac{\partial k_G (y,\epsilon)}{\partial y} & = & \frac{ \alpha_s N_c}{\pi}  \{ C + ln [\kappa_G (x,Q^2)] + E_1 [\kappa_G (x,Q^2)]\} \nonumber \\ 
& \equiv & F(\kappa_G)  \,\,.
\label{agl3}
\end{eqnarray}
In what follows we will present the asymptotic solution of the above equation \cite{ayala1}. Before some comments are necessary: 
\begin{itemize}
\item This equation matches the DLA limit of the DGLAP  evolution equation in the limit of low parton densities $(\kappa \rightarrow 0)$;
\item In first order of unitarity corretions [${\cal{O}}(\kappa^2)$] the AGL equation matches the GLR equation;
\item The AGL equation sums all contributions of the order of $\kappa^n$ absorbing them in the gluon distribution.
\end{itemize}
From the analysis of the AGL equation we get that this equation describes the region of large $\kappa$ (large parton densities) where the  use of  GLR equation is not a good approximation.

We expect that the unitarity corrections start to  be important in kinematical region where $\kappa_G \ge 1$. Assuming $\kappa_G = 1$ and using the definition of $\kappa_G$ [Eq. (\ref{kapag})] we can approximately estimate the behavior of the gluon distribution in the region where the parton densities become large
\begin{eqnarray}
\kappa_G = 1 \,\,\, \Rightarrow \,\,\, xG(x,Q^2) = \frac{2Q^2 \,R^2}{3 \pi \alpha_s}\,\,.
\label{kapaum}
\end{eqnarray}
Therefore in a first approximation the gluon distribution saturates at small values of $x$. Below we will consider the asymptotic regime and demonstrate that this behavior is modified in the region where $k_G > 1$ (very large densities).

Analysing the structure of the Eq. (\ref{agl3}) we see that it has a solution which depends only on $y$. In \cite{ayala1} was shown that  this solution is the asymptotic solution of the AGL equation. In this case we have
 \begin{eqnarray}
\frac{\partial k_G^{asymp} (y,\epsilon)}{\partial y} = F(\kappa_G^{asymp})  \,\,,
\label{aglasy}
\end{eqnarray}
with the solution
\begin{eqnarray}
\int_{ k_G^{asymp} (y = y_0)}^{ k_G^{asymp} (y)} \frac{ d \kappa_G^{\prime}}{F(\kappa_G^{\prime})} = y - y_0 \,\,.
\label{solasy}
\end{eqnarray}
The determination of the behavior of the solution (\ref{solasy}) in the general case  is a difficult task. However, in the limit of high parton densities we have that $ F(\kappa_G^{asymp}) \rightarrow \frac{\alpha_s N_c}{\pi}\, ln\, \kappa_G^{asymp} $. Consequently, 
\begin{eqnarray}
\int_{ k_G^{asymp} (y = y_0)}^{ k_G^{asymp} (y)} \frac{ d \kappa_G^{\prime}}{ln \kappa_G^{\prime}} = \frac{\alpha_s N_c}{\pi}\,\, (y - y_0) \,\,.
\label{solasy2}
\end{eqnarray}
Therefore at large values of densities and $y \gg y_0$  the asymptotic solution is given by
\begin{eqnarray}
 k_G^{asymp} (y) = \frac{\alpha_s N_c}{\pi} \,\, y ln \,y \approx  \frac{\alpha_s N_c}{\pi} \,\, y  \,\,.
\label{kasy}
\end{eqnarray}
We have checked that this solution is a good approximation at very small values of $x$ [${\cal{O}}(10^{-8})$], where we expect that the density of partons will be very large, generating    a high dense system. Moreover,  as the energy behavior of the $\kappa$ function result  from the $y = ln 1/x$ factor, we will assume in this paper that the  asymptotic solution is given by (\ref{kasy}) and use  this expression in the analysis that follows.


Substituting the definition of $\kappa_G$ [Eq. (\ref{kapag})] in the above solution, we  have that in asymptotic regime the behavior of the gluon distribution is given by (for fixed $\alpha_s$)
\begin{eqnarray}
xG(x,Q^2) = \frac{2N_c Q^2 R^2}{3 \pi^2} \,\, ln \,(\frac{1}{x}) \,\,.
\label{gluonasy}
\end{eqnarray}     
Therefore,  the gluon distribution do not saturate at small values of $x$, but is linearly proportional to $ln \, s$ $(s \approx 1/x)$. However, this behavior is softer than predicted by the DGLAP  equation ($xG \propto exp [\sqrt{ ln 1/x}]$)  and the BFKL equation ($xG \propto x^{- \lambda}$, $\lambda > 0$). We have that the gluon distribution present a partial saturation in its behavior. 
Moreover, we have that the gluon distribution is directly dependent of the free parameter $R$, introduced by the profile function. In general this parameter is identified with the proton radius. However, $R$ is associated 
with the spatial gluon distribution within  the proton, which may be smaller than the 
proton radius (see discussion in Ref. \cite{vic2}).

An identical dependence in $Q^2$ and $x$ for the behavior of the gluon distribution in the asymptotic regime was obtained in Ref. \cite{levtu}, where the solution of the K equation in the DLA limit was considered. This result is expected due to the equivalence between the K and AGL approaches in the DLA limit demonstrated in the Ref. \cite{npbvic}.

In Ref. \cite{muesat} Mueller has argued that the factor $ln \, 1/x$ in the gluon distribution is associated with the one loop level of the calculations and that beyond of this level the distribution has the same form of the Weizacker-Williams gluon distribution, which is independent of the energy. However, this statement contrast with the studies of the parton evolution in the non-linear region \cite{ayala1,jamalwang}, where the saturation is not present. The understanding of the difference between these results is an important open question. In this paper we assume (\ref{gluonasy}) as the behavior of the gluon distribution in the asymptotic regime.

In the next section  we consider the consequence of  (\ref{gluonasy})  in the     behavior of the structure function. Before  we will consider the implication of the use of running $\alpha_s$  in the asymptotic solution of the AGL equation and the comparison between the predictions of the  AGL and MV-JKLW approaches in the asymptotic regime.

\subsection{The AGL equation for running $\alpha_s$ }

The solution  (\ref{gluonasy}) was obtained from the AGL equation  [Eq. (\ref{agl3})] in the DLA limit of perturbative QCD and fixed $\alpha_s$. Our goal in this subsection is to shown that although the AGL equation is modified by the running of $\alpha_s$  the general behavior of the gluon distribution, {\it i.e.} the partial saturation, is independent of the approximation used. 

Since the QCD coupling constant is given by $\alpha_s (Q^2) = 4 \pi /(\beta_0 \epsilon)$, where $\beta_0 = 11 - 2/3 n_f$ ($n_f$ is the number of flavors) and $\epsilon = ln \, Q^2/\Lambda_{QCD}^2$,  the relation between the gluon distribution and the function $\kappa_G$ can be expressed by
\begin{eqnarray}
xG(x,Q^2) = \frac{2 Q^2 R^2}{N_c \pi} \frac{\beta_0}{4 \pi}\, \epsilon \, \kappa_G (x,Q^2) \,\,.
\end{eqnarray}
Using this result in  Eq. (\ref{agl2}) we obtain that the AGL equation with running $\alpha_s$ is given by
\begin{eqnarray}
\frac{\partial^2 k_G (y,\epsilon)}{\partial y \partial \epsilon} + \left( \frac{1}{\epsilon} + 1 \right) \frac{\partial k_G (y,\epsilon)}{\partial y} & = & \frac{N_c \alpha_s (Q^2)}{\pi}  \{ C + ln [\kappa_G (x,Q^2)] \nonumber \\
& + & E_1 [\kappa_G (x,Q^2)]\} \nonumber \\
& \equiv & H(\kappa_G) \,\,.
\label{runagl}
\end{eqnarray}
The AGL equation at fixed $\alpha_s$ is a direct consequence of the above equation in the limit of large $\epsilon$.
Moreover, this  equation also has a solution which depends only on $y$. In this case we have
\begin{eqnarray}
\frac{\partial k_G^{asymp} (y,\epsilon)}{\partial y} = \frac{\epsilon}{1+ \epsilon} H(\kappa_G)
\label{asyrun}
\end{eqnarray}
with the solution
\begin{eqnarray}
\int_{ k_G^{asymp} (y = y_0)}^{ k_G^{asymp} (y)} \frac{ d \kappa_G^{\prime}}{H(\kappa_G^{\prime})} = \frac{\epsilon}{1+ \epsilon} \, (y - y_0) \,\,.
\label{solasyrun}
\end{eqnarray}
Following the same steps used in the previous subsection we get that the asymptotic behavior of the gluon distribution is given by
 \begin{eqnarray}
xG(x,Q^2) = \frac{\epsilon}{1+ \epsilon} \,\frac{2N_c Q^2 R^2}{3 \pi^2} \,\, ln \,(\frac{1}{x}) \,\,.
\label{gluonasyrun}
\end{eqnarray}
As expected at large values of $\epsilon$ $(Q^2)$ this solution reduces to the solution for fixed $\alpha_s$. The main difference is the behavior at small values of $\epsilon$, where the prefactor in (\ref{gluonasyrun}) is important. However, the partial saturation of the gluon distribution is not modified by the running  of $\alpha_s$. It agrees with the result that the unitarity corrections are expected to be relevant before of the next to leading order corrections \cite{mueplb}.

\subsection{MV-JKLW approach in the asymptotic regime}

Recently an approach to the evolution of dense partonic systems within the framework of the Wilson renormalization group and the McLerran-Venugopalan approach was developed in Ref. \cite{jamal2}. This approach results in a nonlinear  evolution equation for the  generating functional of the color charge density correlators, which is valid to leading order in $\alpha_s$ at densities which parametrically do not exceed $1/\alpha_s$. The equation  is fairly complicated since it only requires ordering in longitudinal momenta during evolution and puts no constraint in the ordering of transverse momenta  (we refer  \cite{mv,jamal2} for details). However, some special limits were considered: at low densities this equation reduces to the BFKL equation and as consequence to the DLA limit of the DGLAP evolution equation at large values of $Q^2$. 
In Ref. \cite{jamal} the authors  
have considered the DLA limit of the functional evolution equation obtaining the following equation for the impact parameter dependent gluon distribution $xG(x,Q^2,b_t)$
\begin{eqnarray}
\frac{\partial^2 xG(x,Q^2,b_t)}{\partial y \partial \epsilon} =
\frac{N_c(N_c - 1)}{2}\,Q^2\, \left[ 1 - \frac{1}{\kappa} \, exp\,\left(\frac{1}{\kappa}\right) \, E_1 \left(\frac{1}{\kappa}\right)\right]\,\,,
\label{mvdla}
\end{eqnarray}
where 
\begin{eqnarray}
\kappa(x,Q^2,b_t) = \frac{2 \alpha_s}{\pi (N_c - 1) Q^2} \, xG(x,Q^2,b_t)\,\,.
\end{eqnarray}
The general solution and the $b_t$ dependence of this evolution equation was discussed in Ref. \cite{jamalwang}.

At large values of densities (large $\kappa$) we can approximate the Eq. (\ref{mvdla}) by \cite{jamal}
\begin{eqnarray}
\frac{\partial^2 xG(x,Q^2,b_t)}{\partial y \partial \epsilon} =
\frac{N_c(N_c - 1)}{2}\,Q^2\,\,\,.
\label{mvdla2}
\end{eqnarray}
To simplify our analysis we will assume that the $b_t$ dependence of  the gluon distribution can be factorized $xG(x,Q^2,b_t) = xG(x,Q^2) S(b_t)$ and consider a central collision ($b_t = 0$), with $S(0) = 1/ \pi R^2$. Consequently, the evolution equation for the gluon distribution at large densities is given by 
\begin{eqnarray}
\frac{\partial^2 xG(x,Q^2)}{\partial y \partial \epsilon} =
\frac{N_c(N_c - 1) \pi}{2}\,R^2 \,Q^2\,\,\,.
\label{mvdla3}
\end{eqnarray}
The solution of this equation is straithforward
\begin{eqnarray}
xG(x,Q^2) = \frac{N_c(N_c - 1) \pi}{2}\,R^2 \,Q^2\, ln \left(\frac{1}{x}\right)\,\,.
\label{mvdlasol}
\end{eqnarray}
Therefore the solution of the Eq. (\ref{mvdla}) has the same $Q^2$ and $x$ dependence  of the AGL equation in the asymptotic regime. The difference in the prefactors is a function of the distinct normalizations and approximations used in the two approaches. The demonstration of the equivalence between the two approaches in all kinematical region is still an open question.

The main conclusion of our analysis  of these two approaches is the universality of the $x$ and $Q^2$ dependence of the gluon distribution (the partial saturation) in the asymptotic regime of the high density QCD. In the next section we will consider the consequences of the partial saturation in the behavior of the  structure function, what is experimentally measured.

\section{The Asymptotic Behavior of the Structure Function}

The common feature of the BFKL and DGLAP equations  is the steep increase  of the 
cross sections as $x$ decreases. This steep increase cannot persist down to arbitrary 
low values of $x$ since it violates a fundamental principle of quantum theory, {\it i.e.} the 
unitarity. In the context of relativistic quantum  field theory of the strong interactions, unitarity 
implies  that the cross section  cannot increase with increasing 
energy $s$ above $log^2 \,s$: the Froissart's theorem \cite{froi}. In this section we will calculate the structure function in the asymptotic regime using  as input the solution of the AGL equation [Eq. (\ref{gluonasy})] which considers the  unitarity corrections.

Lets us start from the space-time picture of the $ep$ processes \cite{gribov}.
The deep inelastic scattering $ep \rightarrow e + X$ is characterized by a large electron energy loss $\nu$ (in the target rest frame) and an invariant momentum
transfer  $q^2 \equiv - Q^2$ between the incoming and outgoing electron 
such that $x = Q^2/2m_N \nu$ is fixed.  The general features of the time 
development  can be established using only
Lorentz invariance and the uncertainty principle. The incoming physical
electron state can, at a given instant of time, be expanded in terms of its
(bare) Fock states
\begin{eqnarray}
|e>_{phys} = \psi_e |e> + \psi_{e\gamma} |e \gamma> + ... \,\,.
\end{eqnarray}
The amplitudes $\psi_i$ depend on the kinematic variables describing the 
states $|i>$, and have the time dependence $exp(-iE_it)$, where $E_i = \sum_i 
\sqrt{m_i^2 + \vec{p}_i^{\,2}}$ is the energy of the state. The 'lifetime'
$\tau_i \approx 1/(E_i - E_e)$ of a Fock state $|i>$ is given by the time
interval after which the relative phase $exp[-i(E_i - E_e)]$ is significantly
different from unity. If $\tau_i > R_A$ the Fock state forms long before the 
electron arrives at the nucleon, and it lives long after its passage. 
New Fock states are not formed inside the nucleon. Therefore,
the scattering inside the nucleon is diagonal in the Fock basis. If the 
state $|i>$ contains particles with mass $m_j$, energy fraction $x_j$ and transverse
momentum $p_{t\,j}$, we have that the transverse velocities $v_{t\,j} =
p_{t\,j}/x_j E_e$ are small at large $E_e$. Hence the impact parameters (transverse
coordinates) of all particles are preserved. 

In terms of Fock states we then view the $ep$ scattering as follows: the 
electron emits a photon ($|e> \rightarrow |e\gamma>$) with $E_{\gamma} = \nu$
and $p_{t \, \gamma}^2 \approx Q^2$, after  the photon splits  into a $q\overline{q}$
($|e\gamma> \rightarrow |e q\overline{q}>$) and typically travels a  
distance $l_c \approx 1/m_N x$, referred as the 'coherence length', 
before interacting in the nucleon. For small $x$, the photon converts to a quark pair at a large distance before it interacts to the target; for example, at the $ep$ HERA collider, where one can study structure functions at $x \approx 10^{-5}$, the coherence length is as large as $10^4 \,fm$, much larger than the nucleon radii.
Consequently, the space-time picture of the DIS in the target
rest frame can be viewed as the decay of the virtual photon at high energy
(small $x$) into a quark-antiquark pair long before the 
interaction with the target. The $q\overline{q}$ pair subsequently interacts 
with the target.  In the small $x$ region, where 
$x \ll \frac{1}{2mR}$, the $q\overline{q}$  pair 
crosses the target with fixed
transverse distance $r_t$ between the quarks. It allows to factorize the total 
cross section between the wave function of the photon and the interaction 
cross section of the quark-antiquark pair with the target. The photon wave function 
is calculable and the interaction cross section is modelled. Therefore, the 
 structure function is given by \cite{nik}
\begin{eqnarray}
F_2(x,Q^2) = \frac{Q^2}{4 \pi \alpha_{em}} \int dz \int \frac{d^2r_t}{\pi} |\Psi(z,r_t)|^2 \, \sigma^{q\overline{q}}(z,r_t)\,\,,
\label{f2target}
\end{eqnarray}
where 
\begin{eqnarray}
|\Psi(z,r_t)|^2 = \frac{6 \alpha_{em}}{(2 \pi)^2} \sum^{n_f}_i e_i^2 \{[z^2 
+ (1-z)^2] \epsilon^2\, K_1(\epsilon r_t)^2 + m_i^2\, K_0(\epsilon r_t)^2\}\,\,,
\label{wave}
\end{eqnarray}
$\alpha_{em}$ is the electromagnetic coupling constant,
$\epsilon^2 = z(1-z)Q^2 + m_i^2$, $m_i$ is the quark mass, $n_f$ is the number 
of active flavors, $e_f^2$ is the square of the  parton charge (in units of $e$), $K_{0,1}$ 
are the modified Bessel functions and $z$ is the fraction of the photon's light-cone 
momentum carried by one of the quarks of the pair.  In the 
leading log$(1/x)$ approximation we can neglect the change of $z$ during the 
interaction and describe the cross section $\sigma^{q\overline{q}}(z,4/r_t^2)$ as 
a function of the variable $x$. Considering only light quarks ($i=u,\,d,\,s$)   $F_2$ 
can be expressed by \cite{plb}
\begin{eqnarray}
F_2(x,Q^2) = \frac{1}{4 \pi^3} \sum_{u,d,s}  e_i^2 \int_{\frac{1}
{Q^2}}^{\frac{1}{Q_0^2}} \frac{ d^2r_t}{\pi r_t^4}\,\sigma^{q\overline{q}}(x,r_t) \,\,.
\label{f2sim}
\end{eqnarray}
We have introduced a cutoff in the superior limit of the integration, in order  to eliminate 
the long distance (non-perturbative) contribution in our calculations.

Using the expression of  $\sigma^{q\overline{q}}$ \cite{plb} [Eq. \ref{sigqq}], we get
\begin{eqnarray}
F_2(x,Q^2) =  \frac{2 \alpha_s}{9 \pi} \int_{Q_0^2}^{Q^2} \frac{d Q^2}{Q^2} \, xG(x,Q^2) \,\,.
\label{f2vio}
\end{eqnarray}
This equation is a leading twist relation which  will eventually break down when we consider the higher twist terms in the evolution. Below we discuss a generalization of the relation between the structure function and the gluon distribution and the implications in the asymptotic behavior. However, before we will assume that the relation (\ref{f2vio}) is valid and will determine the behavior of the structure function in the asymptotic regime.

Using the solution of the AGL equation in the asymptotic regime [Eq. (\ref{gluonasy})] as input in the Eq. (\ref{f2vio}) we get 
\begin{eqnarray}
F_2(x,Q^2) \approx  \frac{ \alpha_s}{ \pi^3}\,R^2\,Q^2 \,ln \left(\frac{1}{x}\right)  \,\,.
\label{f2asy}
\end{eqnarray}
 We see that the partial saturation of the gluon distribution implies that the $F_2$ structure function does not saturate at small values of $x$, but is linearly proportional to  $ln \, s$. The main point is that this behavior does not violate the Froissart boundary. 

An identical result for the behavior of the structure function was obtained by Kovchegov in  Ref. \cite{kov2}, using as input  the solution of the  evolution equation proposed in \cite{kov}.  This evolution equation (the K equation) considers the multiple pomeron exchanges in the leading logarithmic $1/x$ approximation [LLA($1/x$)], which implies the unitarization of BFKL Pomeron. In Ref. \cite{npbvic} we have shown that the 
K equation is equivalent to the AGL equation in the DLA limit of perturbative QCD.
The equivalence between the asymptotic behavior of the structure function obtained from the AGL approach (DLA) and the K approach [LLA($1/x$)] shows that the  asymptotic behavior (28) is a general characteristic which is independent of the approximations used in the calculations.  Furthermore, as the predictions of the AGL and MV-JKLW approaches for the gluon distribution are identical in the asymptotical regime, we have that the asymptotic  behavior of the $F_2$ structure function is well-established by the three distinct approaches.

A comment is in order here. If we assume that the behavior of the gluon distribution is given by (\ref{kapaum}), which is valid in the region where the density starts to becomes large, we then get that the structure function saturates at small values of $x$ ($F_2 \propto R^2 Q^2$). However, as shown in the previous discussion, this result is valid in the  limited kinematical region where the higher orders in the parton density are not important. For a longer theoretical and phenomenological discussion about the parton saturation see for example the Refs. \cite{muesat,wust}.

We now consider that the relation between the structure function and the gluon distribution is modified by the particular type of higher twist terms associated with the unitarity corrections (See also \cite{mv2} for a similar calculation in the infinite momentum frame). First we obtain the expression for the structure function considering these corrections and  after we estimate the behavior of $F_2$ in the asymptotic regime.

We estimated the unitarity corrections considering the  $s$-channel unitarity constraint in the interaction cross section of the quark-antiquark pair with the target \cite{ayala1}. In this case we  the structure function is given by 
\begin{eqnarray}
F_2(x,Q^2) =  \frac{1}{2\pi^3} \sum_{u,d,s} e_f^2 \int_{\frac{1}{Q^2}}^{\frac{1}{Q_0^2}} \frac{d^2r_t}{\pi r_t^4} \int d^2b_t 
\{1 - e^{-\frac{1}{2}\sigma^{q\overline{q}}(x,4/r_t^2)S(b_t)}\}\,\,.
\label{f2eik}
\end{eqnarray}

The use of the Gaussian parametrization for
the  nucleon profile function $S(b_t) = \frac{1}{\pi R^2} e^{-\frac{b^2}{R^2}}$, 
where $R$ is a free parameter, as before,   simplifies the calculations.
We obtain that   
the $F_2$ structure function can be written  as \cite{ayala1}
\begin{eqnarray}
F_2(x,Q^2) =  \frac{R^2}{2\pi^2} \sum_{u,d,s} \epsilon_i^2 \int_{\frac{1}{Q^2}}^{\frac{1}{Q_0^2}} \frac{d^2r_t}{\pi r_t^4} \{C + ln(\kappa_q(x, r_t^2)) + E_1(\kappa_q(x, r_t^2))\}\,\,,
\label{diseik2}
\end{eqnarray}
where  $\kappa_q = 4/9 \kappa_G =  (2 \alpha_s/3R^2)\,\pi\,r_t^2\,
 xG_N(x,\frac{1}{r_t^2})$.
This equation allows to estimate the unitarity corrections to the  structure function  in the DLA limit.
 Expanding the  equation (\ref{diseik2})  for small $\kappa_q$, 
the first term (Born term) will correspond to the usual DGLAP equation in the small $x$ 
region [Eq. (\ref{f2vio})].

In the asymptotic regime (large $\kappa_q$) we obtain 
\begin{eqnarray}
F_2(x,Q^2) \approx  \frac{R^2}{2\pi^2} \sum_{u,d,s} \epsilon_i^2 \int_{\frac{1}{Q^2}}^{\frac{1}{Q_0^2}} \frac{d^2r_t}{\pi r_t^4} \,ln(\kappa_q(x, r_t^2)) \,\,.
\label{diseikasy}
\end{eqnarray}
Using the asymptotic solution of the AGL equation we can determine $\kappa_q$ at large values of densities, and so 
\begin{eqnarray}
F_2(x,Q^2) \approx  \frac{R^2}{2\pi^2} \sum_{u,d,s} \epsilon_i^2 \int_{\frac{1}{Q^2}}^{\frac{1}{Q_0^2}} \frac{d^2r_t}{\pi r_t^4} \,ln\,\left[
\frac{4 \alpha_s}{3} \, ln \, (\frac{1}{x})\right] \,\,.
\label{diseikasy2}
\end{eqnarray}
Therefore, considering the contribution of the  higher twist terms in the relation between the structure function and the gluon distribution, we predict the following asymptotic behavior for the structure function
\begin{eqnarray}
F_2(x,Q^2) \approx  \frac{R^2 Q^2}{3 \pi^2}  \,ln\,\left[
\frac{4 \alpha_s}{3} \, ln \, (\frac{1}{x})\right] \,\,.
\label{diseikasy3}
\end{eqnarray}
We see that the inclusion of higher twist term implies a softer dependence of $F_2$ with the energy than obtained using the leading twist relation. However, in both cases the structure function does not violate the Froissart boundary in the asymptotic regime of high density QCD. The demonstration of the behavior (\ref{diseikasy3}) using the other approaches for hdQCD is still an open question.

From the results for the structure function in the asymptotic regime we can see that this regime is characterized by the identity
\begin{eqnarray}
\frac{d F_2(x,Q^2)}{ d ln \, Q^2} = F_2 (x,Q^2)\,\,,
\end{eqnarray}
which is an important signature of the asymptotic regime of high density QCD.
This regime should be reached for the case of an interaction with nuclei at smaller parton densities than in a nucleon, since $\kappa_A = A^{\frac{1}{3}} \times \kappa_N$, where $\kappa_N$ is given by the expression (\ref{kapag}).

\section{Conclusion}

In this paper we have analysed the asymptotic regime of high density QCD. We have shown that the partial saturation of the gluon distribution is a well-established result from the AGL equation for running and fixed $\alpha_s$ and from the MV-JKLW approach in this kinematical regime. Using the asymptotic behavior of the gluon distribution we calculate the $F_2$ structure function assuming  first that the leading twist relation between these two quantities is valid and second that this relation is modified by the higher twist terms associated to the unitarity corrections. In the first case we have obtained  that the corresponding $F_2$ structure function  is linearly proportional to $ln \, s$, which agrees with the results obtained recently  by Kovchegov using a distinct approach. In the second case a softer behavior is obtained. In both cases, the $F_2$ structure function unitarizes and the Froissart boundary is not violated in the asymptotic regime of high density QCD. We conclude that the partial saturation of the gluon distribution and the unitarization of the structure  function   are general results, established in three distinct approaches for high dense QCD systems. However, the demonstration of the equivalence between the distinct approaches in all kinematical regions and of the generality of the behavior of $F_2$ when the higher twist are considered  are still important open questions.

\section*{Acknowledgments}

This work was partially financed by CNPq and by Programa de Apoio a N\'ucleos de Excel\^encia (PRONEX), BRAZIL.

\end{document}